\documentclass[english,graybox]{svmult}
\usepackage{mathptmx}
\usepackage{helvet}
\usepackage{courier}
\usepackage[T1]{fontenc}
\usepackage[latin9]{inputenc}
\usepackage{verbatim}
\usepackage{float}
\usepackage{url}
\usepackage{amsmath}
\usepackage{graphicx}
\usepackage{txfonts}
\usepackage[round]{natbib}

\makeatletter

\renewcommand{\cite}{\citet}

\usepackage{type1cm}

\usepackage{multicol} 
\usepackage[bottom]{footmisc} 

\makeatother

\usepackage{babel}

\def\kms{km ${\rm s}^{-1}$}
\def\sn1{${\rm s}^{-1}$}
\def\cmd{${\rm cm}^{-3}$}
\def\cmn{${\rm cm}^{-2}$}
\def\ergs{erg ${\rm s}^{-1}$}

\def\chan{{\it Chandra}}

\def\apm{APM 08279+5255}
\def\sp{\space}

\def\ngc3783{NGC~3783}
\def\mr{MR~2251$-$178}

\def\logxi1{erg cm ${\rm s}^{-1}$}

\def\xstar{{\sc xstar}}
\def\cloudy{{\sc cloudy}}
\def\rmxaa{Rev. Mexicana Astron. Astrofis.}%
\def\apj{Astrophysical Journal}%
\def\apjl{Astrophysical Journal Letters}%
\def\apjs{Astrophysical Journal S.}%
\def\aap{A\&A}%
\def\ssr{Space Sci. Rev.}%

\begin{document}

\title*{X-ray outflows of active galactic nuclei warm absorbers: A 900 ks Chandra simulated spectrum}

\titlerunning{X-ray outflows of AGNs}

\author{J.M Ram{\'i}rez-Velasquez and Javier Garc{\'i}a}

\authorrunning{Ram{\'i}rez-Velasquez and Garc{\'i}a}

\institute{J.M Ram{\'i}rez-Velasquez \at Physics Centre, Venezuelan Institute for Scientific Research ({\sc ivic}), PO Box 20632, Caracas 1020A, Venezuela - 
\email{josem@ivic.gob.ve}
and
Departamento de Matem\'aticas, {\sc cinvestav} del I.P.N.,
07360 M\'exico, D.F., M\'exico.
\and
Javier Garc{\'i}a \at 
Harvard-Smithsonian Center for Astrophysics
60 Garden Street, Cambridge, MA 02138, USA - \email{javier@head.cfa.harvard.edu}
}

\maketitle

\abstract{
            We report on the performance of the statistical,
            X-ray absorption lines identification 
            procedure {\sc xline-id}.
            As illustration,
            it is used to estimate the
            time averaged gas density $n_H(r)$
            of a representative AGN's warm absorber ($T\approx 10^5$~K)
            X-ray simulated spectrum.
            The method relies on three
            key ingredients: (1) a well established emission continuum
            level; (2) a robust grid of photoionization
            models spanning several orders of magnitude in gas density ($n_H$),
            plasma column density
            ($N_H$), and
            in ionization states; (3) theoretical curves of growth for a large
            set of atomic lines. By comparing theoretical and observed
            equivalent widths of a
            large set of lines, spanning
            highly ionized charge states
            from O, Ne, Mg, Si, S, Ar, and the Fe L-shell
            and K-shell,
            we are able to
            infer the location of the X-ray
            warm absorber.
}

\section{Introduction} 

In active galactic nuclei (AGNs),
photoionization has been recognized as the main mechanism
for the formation of  Ultraviolet (UV) and X-ray lines
(\cite{bahcall1969a}).
If the photo-ionized plasma reaches a stable state for their
ion populations, they are said to be in photo-ionization
equilibrium (PIE). Under the assumption of PIE, one of the
principal parameters describing the physical condition
of the plasma is the ionization parameter $\xi$:
\begin{equation}
\xi(r)=\frac{L}{n_H(r)r^2},
\label{xi11}
\end{equation}
where $L$ is the ionizing luminosity of the primary source,
$n_H(r)$ is the gas density,
and $r$ is the distance from the ionizing source to the
interacting shell/slab of plasma. Therefore, spectral fitting
to the observational data in the X-ray band of a given source
combined with a measurement of its luminosity
$L$ provide constrains on the degenerated quantity $n_H(r)r^2$.
Moreover, in order to have an estimate of the spatial
location of the gas absorbing/emitting photons from the primary
underlying source, an independent determination of the gas
density $n_H(r)$ is required.

Curve of growth (COG) (\cite{spitzer1998a}) is a useful
tool to gather information from astrophysical spectra, like
ion column density ($N_{ion}$) using the strength of
absorption lines. However, the technique is full of 
exact requirements and limitations. In order for a line
to be used as COG diagnostic tool,
firstly (1) a well defined continuum level has to be known
prior any construction of the COG.
Secondly (2) the synthetic COGs, are usually compare with real
measurements of equivalent widths (EWs) of lines.
Third (3) the chosen lines must be not saturated
respect to the variable we wish to contrast to.
Furthermore (4), a reliable {\it  identification}
procedure {\it must be} built on the grounds of a well constructed
photoionization modelling (\cite{kallman2001a})
which in turn needs a well
established set of atomic data.
For instance, \cite{badnell2006a}
pointed out about the inappropriate use of some dielectronic
recombination (DR) rates, which are later discussed by
\cite{kallman2010a} in the context of photoionization modelling
and the correct use of atomic data (now implemented in the latest
version of \xstar)
(\cite{bautista2001a}).

Following a series of papers related with the computations
of K lines of Fe (\cite{palmeri2002a,palmeri2003a}),
and K-shell photoabsorption of O ions (\cite{ramirez2002a,garcia2005a,garcia2011b}
atomic data (energies, cross sections, lifetimes),
\cite{palmeri2008a}) discuss in detail the reliability and accuracy
of some sets of these data (medium-$Z$ elements),
currently observed in the X-ray spectrum of
active galactic nuclei 
(AGNs, e.g., \ngc3783 \sp \cite{kaspi2002a,ramirez2005a,ramirez2011a},
\mr \sp \cite{ramirez2008b}, \apm \sp \cite{ramirez2008a}, Ark 564
\cite{ramirez2013a}),
in the form of K-absorption lines of H- and He-like
Ne,Mg,Si,S, possibly Ar and Ca, and also from lower and medium
ionization stages of Si and S
(e.g., for MCG-6-30-15 and IRAS 13349+2438)
(\cite{holczer2007a}),
with the aim of improving the atomic database of the
\xstar \sp modelling code.

We report on the accuracy of the statistical
method, {\sc xline-id}
which allows us to extract 
time averaged
gas density
$n_H$ of interacting material surrounding
UV+X-rays sources,
from EW measurements and {\it detection}
of unsaturated
absorption lines respect $n_H$.

The paper is organized as follows:
in the first section, the set of data used,
next the details of the method, and
finally
we delineate the results of the distribution
of Doppler velocities found, and compute gas densities.
We also discuss the results and conclude.
Throughout this paper, we use a cosmology with
H$_0=70$ \kms \space Mpc$^{-1}$,
$\Omega_M=0.3$ and $\Omega_{\lambda}=0.7$.

\begin{figure}
\centering
\includegraphics[angle=-90,width=10.5cm]{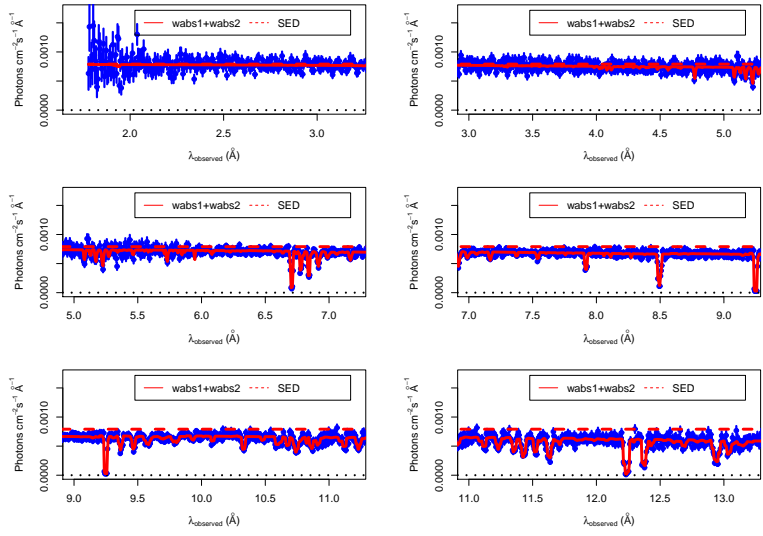}
\caption{The continuum emission seen by the material surrounding the UV+X-rays source
($1-13$ \r{A}). Points with error bars are the 900 ks simulated redshifted {\it Chandra} observation.
Solid lines: photoionization model of the absorber (red), and dashed line: the SED.}  
\label{conti_no_abs1}
\end{figure}

\begin{figure}
\centering
\includegraphics[angle=-90,width=10.5cm]{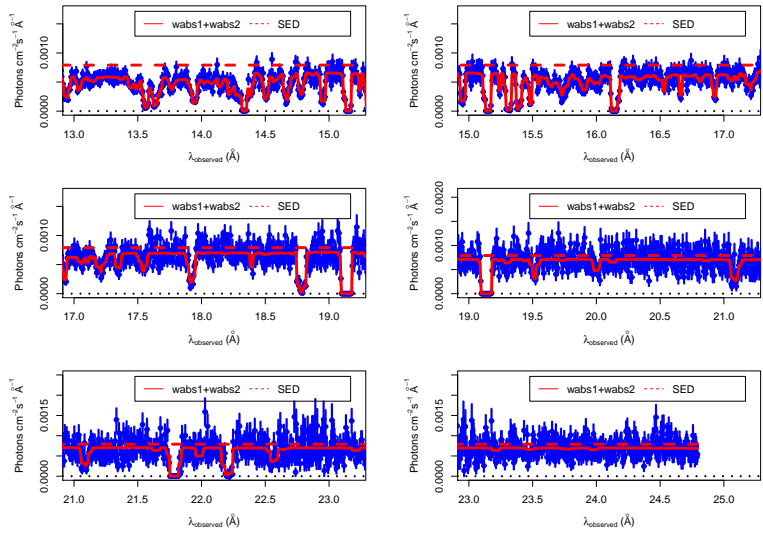}
\caption{The continuum emission seen by the material surrounding the UV+X-rays source
($13-25$ \r{A}). Points with error bars are the 900 ks simulated redshifted {\it Chandra} observation.
Solid lines: photoionization model of the absorber (red), and dashed line: the SED.}
\label{conti_no_abs2}
\end{figure}

\section*{The set of data and the emission continuum level
\label{datainput}}

One of the main problems in the identification of narrow spectral
features in X-ray sources, is 
the impossibility of contrasting the final atomic identification (ID)
of a feature with an
input feature, once it has passed through all the
instrumentation (effective areas, re-distribution matrix, etc).
This is why we aim at assessing this problem, generating
a spectrum, over which we have absolute control of every
single feature (absorption lines and edges).
For that purpose we have created an artificial
first dispersion order
$\pm 1$ HETG+MEG 900 ks
(HETGS with medium energy grating [MEG]), 
where we know exactly where
every feature is located, and described in more detail as follows.


We begin by building a physical model
{\tt (pow)*wabs1*wabs2}
constituted by an UV+X-rays source,
redshifted by $z=0.00976$, emitting
a powerlaw ({\tt pow}),
modified by two
constant-density ($\log[n_H]=12$ [\cmd])
absorbers
({\tt wabs1}, 
with 
ionization parameter $\log \xi=2.08$ [\logxi1],
column densities $N_H=10^{22}$ \cmn,
and
{\tt wabs2},
with
ionization parameter $\log \xi=1.15$ [\logxi1],
column densities $N_H=10^{21}$ \cmn)
\footnote{
Computed using the code \xstar \sp and described in more detail in
the section {\bf Models: The grid}}.
We call this model A.
We then use {\sc xspec} 12 to generate a
900 ks MEG spectrum, using appropriate
+1 and -1 dispersion orders of MEG effective areas
and 1$^{\rm st}$ order response matrix
(see science threads in the {\sc ciao} main page,
{\tt http://cxc.harvard.edu/ciao/}).

Table \ref{tbl1} gives the physical
parameters of model A
to produce one of
the spectra we use in the analysis.
Figures (\ref{conti_no_abs1}-\ref{conti_no_abs2})
show: the simulated $\pm 1$ MEG spectrum,
(points with error bars);
how {\tt wabs1} and {\tt wabs2} (solid red line)
modify the 
Spectral Energy Distribution (SED)\footnote{
We refer also as the
unabsorbed emission continuum from the primary source
as seen by the absorber (dashed line in the figures).}
{\tt pow}.

\begin{table*}
 \centering
 \begin{minipage}{100mm}
  \caption{Physical parameters of model A.}
  \label{tbl1}
  \begin{tabular}{@{}lll@{}}
  \hline
  {\tt Component}&
  {\tt Value}&
  {\tt Units}\\
  \hline

$\Gamma_{\rm x}$                    &  2                                &  --               \\
$N^{\rm a}$                         &  $0.010$                          &  photons keV$^{-1}$cm$^{-2}$s$^{-1}$ at 1 keV                 \\
$z_{\rm pow}$                       &  $0.00976$                        &  -- \\
$\log \xi(1) $                      &  2.08                             & erg cm ${\rm s}^{-1}$   \\
$N_{\rm H}(1)$                      &  $10^{22}$                        & cm$^{-2}$               \\
$z_{\rm out-wabs1}$                 &  $0.00876$                        & --       \\
$\log \xi(2) $                      &  1.15                             & erg cm ${\rm s}^{-1}$   \\
$N_{\rm H}(2)$                      &  $10^{21}$                        & cm$^{-2}$               \\
$z_{\rm out-wabs2}$                 &  $0.00836$                        & --       \\
\hline
\end{tabular}\\
\end{minipage}
\end{table*}

\begin{table*}
 \centering
 \begin{minipage}{100mm}
  \caption{Physical parameters of model B.}
  \label{tbl1c}
  \begin{tabular}{@{}lll@{}}
  \hline
  {\tt Component}&
  {\tt Value}&
  {\tt Units}\\
  \hline


{\tt ism}                           &  $9.91\times 10^{20}$             &   cm$^{-2}$       \\

$\Gamma_{\rm x}$                    &  2                                &  --               \\
$N^{\rm a}$                         &  $0.010$                          &  photons keV$^{-1}$cm$^{-2}$s$^{-1}$ at 1 keV                 \\
$z_{\rm pow}$                       &  $0.00976$                        &  -- \\
$\log \xi(1) $                      &  2.08                             & erg cm ${\rm s}^{-1}$   \\
$N_{\rm H}(1)$                      &  $10^{22}$                        & cm$^{-2}$               \\
$z_{\rm out-wabs1}$                 &  $0.00876$                        & --       \\
$\log \xi(2) $                      &  1.15                             & erg cm ${\rm s}^{-1}$   \\
$N_{\rm H}(2)$                      &  $10^{21}$                        & cm$^{-2}$               \\
$z_{\rm out-wabs2}$                 &  $0.00836$                        & --       \\
\hline
\end{tabular}\\
\end{minipage}
\end{table*}

In order to analyze an X-ray spectrum more similar to a real observation
(\cite{ramirez2013b}),
we introduce an outflow velocity to each
absorber,
$z_{\rm out-wabs1}=0.00876$ (300 \kms) for wabs1, and
$z_{\rm out-wabs2}=0.00836$ (420 \kms) to wabs2.

Finally, one additional complication is introduced in model B.
We add absorption due to interstellar medium (ism),
{\tt ism*(pow)*wabs1*wabs2}, so we can study the effect of measuring
equivalent widths, with and without consider the unabsorbed
continuum emission.
Table \ref{tbl1c}
give parameters for model B.
Each case will be carefully analyzed in the next sections.

\section*{The {\sc xline-id} method: General Picture
\label{cogs1}}

Our goal is to compare measured line equivalent widths (EWs) with
synthetic curves generated as functions of the main physical
parameters of the system:
column density in the line of sight ($N_H$),
(log of) ionization parameter ($\log \xi$),
and
gas density ($n_H$).

The first (1) step to achieve this is to generate a grid of photoionization
models (as in AGNs, photoionization is the dominant mechanism for production
of spectral lines).
Secondly (2), we define the line profile and compute the integrals involved
in the EWs of the lines of interest 
(a list of {\it detected} features is required as input for {\sc xline-id}).
This is a critical step, since none of the main photoionization codes
(e.g., \cloudy, \xstar), compute this quantity (by default) 
for the thousands of lines usually
included in the run of the grids, mainly for computational efficiency.
The third (3) step is to compare measured EWs with theoretical EWs in order
to draw out the parameters of interest.

\subsection*{Models: The grid
\label{grids1}}

Here we describe the building of the grid of models. We use the photoionization
code \xstar \footnote{\tt https://heasarc.gsfc.nasa.gov/docs/software/xstar/xstar.html}
v 2.2 with
the most up-to-date atomic database v 2.2.1bn20 (\cite{bautista2001a}).

The code includes all the relevant
atomic processes,
including inner shell processes 
(\cite{palmeri2003a,palmeri2003b}).
It computes the emissivities
and optical depths
of the most prominent X-ray and UV lines identified in AGN spectra.
Our models are based on spherical slabs
illuminated by a point-like X-ray continuum source.
The input parameters are the source spectrum, the gas
composition, the gas density $n_{H}$, 
the column density in the line of sight
$N_{H}$
and
the ionization parameter
$\xi$. 
The source spectrum is described by the
spectral luminosity
$L_{\epsilon}=L_{\rm ion}f_{\epsilon}$, where $L_{\rm ion}$ is the
integrated
luminosity from 1 to 1000 Ryd, and $\int_{1}^{\rm 1000~Ryd} f_{\epsilon}
d\epsilon=1$.
The spectral function $f_{\epsilon}$ is taken to be 
a powerlaw $\sim \epsilon^{\alpha}$
with $\alpha=-1$ (photon index $\Gamma_{\rm x}=2$).
The gas consists
of the following elements, H, He, C, N, O, Ne, Na, Mg, Si, S,
Ar, Ca and Fe. We use the abundances of \cite{grevesse1996a}
in all our models (and the term {\it solar} for these abundances).
We adopt a turbulent velocity of 200 \kms.
The total ionizing luminosity used is $L_{\rm ion}=2.5 \times 10^{44}$
\ergs.

We have taken two representative ionization parameters:
$\log \xi =2.08$ [\logxi1], and $\log \xi =1.15$ [\logxi1].
We call them
medium ionization plasma (MIP, represented by wabs1)
and
low ionization plasma (LIP, represented by wabs2),
respectively.
Finally due to our interest in testing the sensitivity of
the line strengths with the gas density, we run
$n_{H}=10^{[5,6,7,8,9,10,11,12,13,14,15,16,17,18]}$ \cmd, and
$N_H=$[1,0.1]$\times 10^{22}$ \cmn,
for each of the
two ionization states, resulting in 28 points in the grid,
making it a modest resolution 
density-COG grid for warm absorber
in AGNs.

\subsection*{{\sc trunk-1}: The identification list model
\label{trunk1}}

We have developed a simple algorithm to identify observed absorption spectral
features. The identification is performed by comparing both the line wavelength
and optical depth with those computed theoretically in our grid of models. For
each model we choose the 500 strongest features, which results in a total
of 14,000 theoretical absorption lines (500 $\times$ 14$n_H$ $\times$ 2$N_H$).
We shall refer to this as our {\it identification list}.
The simplest way to identify an observed line is to compare its measured
wavelength with the theoretical values. However, one must also allow for the
possibility of wavelength shifts due to velocity components in the absorbing
material. Thus, starting with the best-fit for the line centroid $\lambda_0$
of a given observed line, we search for theoretical lines within a
$\pm\Delta\lambda_0$ identification window (IW) around $\lambda_0$. This allows for
the identification of blue or redshifted lines. If more than one theoretical
line is present in the identification window, we choose the line with the
largest optical depth at the core ($\tau_0$). 

A second observed feature can be
found close enough such that its identification window cover a similar range,
including the same strong lines. If the next observed feature is identified by
the same theoretical line as in the feature just before, we do not allow for
repetition but rather choose the second strongest line in the identification
window. This allows for the detection of duplets. Nevertheless, this constrain
is only applied to two consecutive lines. Thus, if a third consecutive observed
line (i.e. two features next) is identify with the same strongest theoretical
lines as the first, this one is assigned again. This allows the detection of
more than one velocity component. The result is an identification list based
on the strongest predicted absorption line, $\pm\Delta\lambda_0$ \AA \sp around
a feature detected at $\lambda_0$ \AA, truncated only {\it once} for repetition
if two features are too close. We call this identification list model {\sc
trunk-1}.

Similar studies have been carried out in the past where measured absorption
features are compared with a theoretical features in a given identification
list (e.g., \cite{krongold2003a,netzer2003a}). However, these studies present
two important limitations: the identification list employed typically contain
a relatively small number of theoretical features (e.g., $\sim 50$), 
and their identification
is based on comparing the wavelength position of the observed feature with
those calculated theoretically
using undefined identification window (how much is allowed for an observed feature
to be compared with theory, e.g., 0.1 \AA, is too large).
The latter restriction is particularly important
because the larger the identification window becomes, the more are the chances
of an incorrect identification. 
\begin{figure}
\centering
\includegraphics[angle=-90,width=8.5cm]{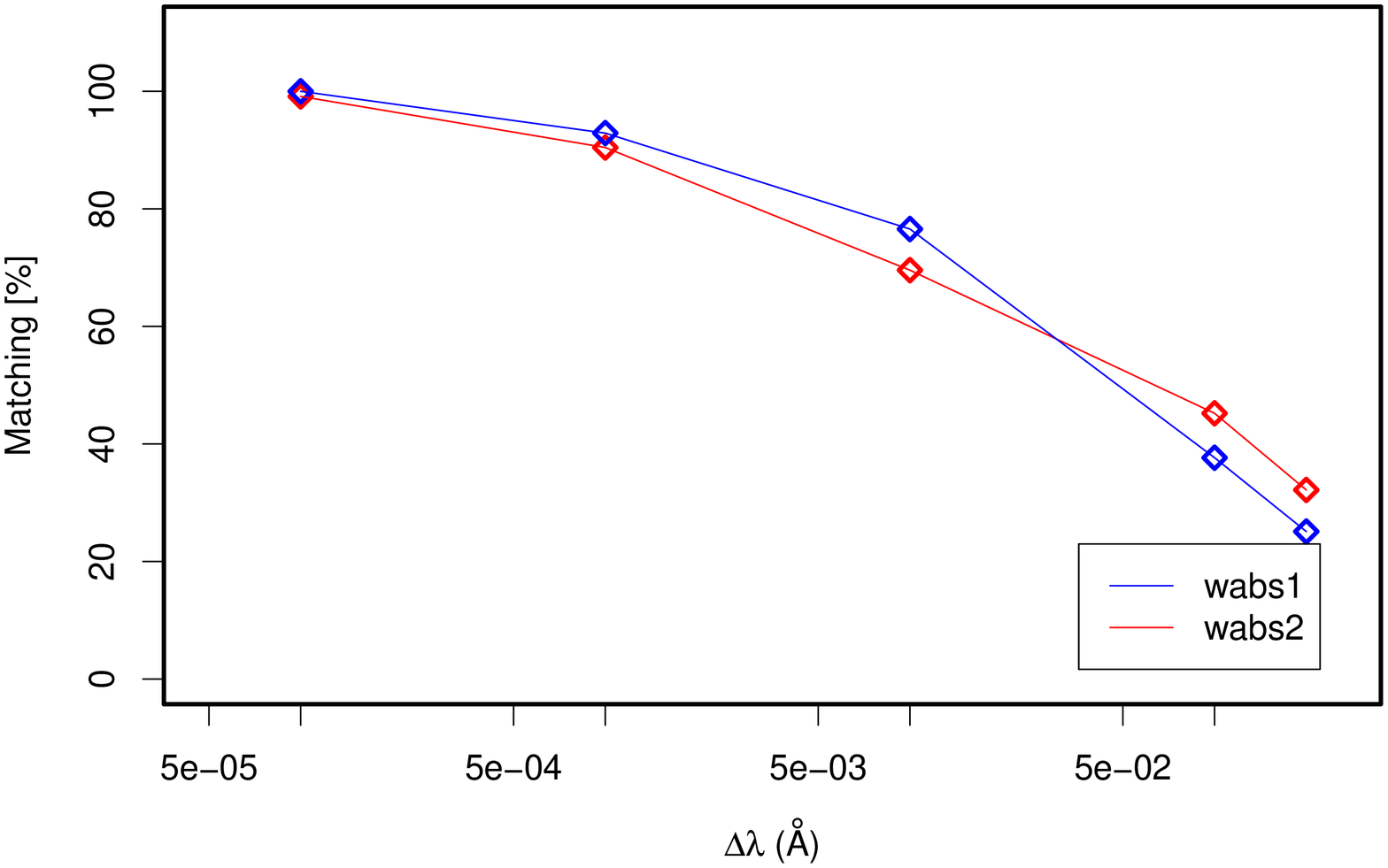}
\includegraphics[angle=-90,width=8.5cm]{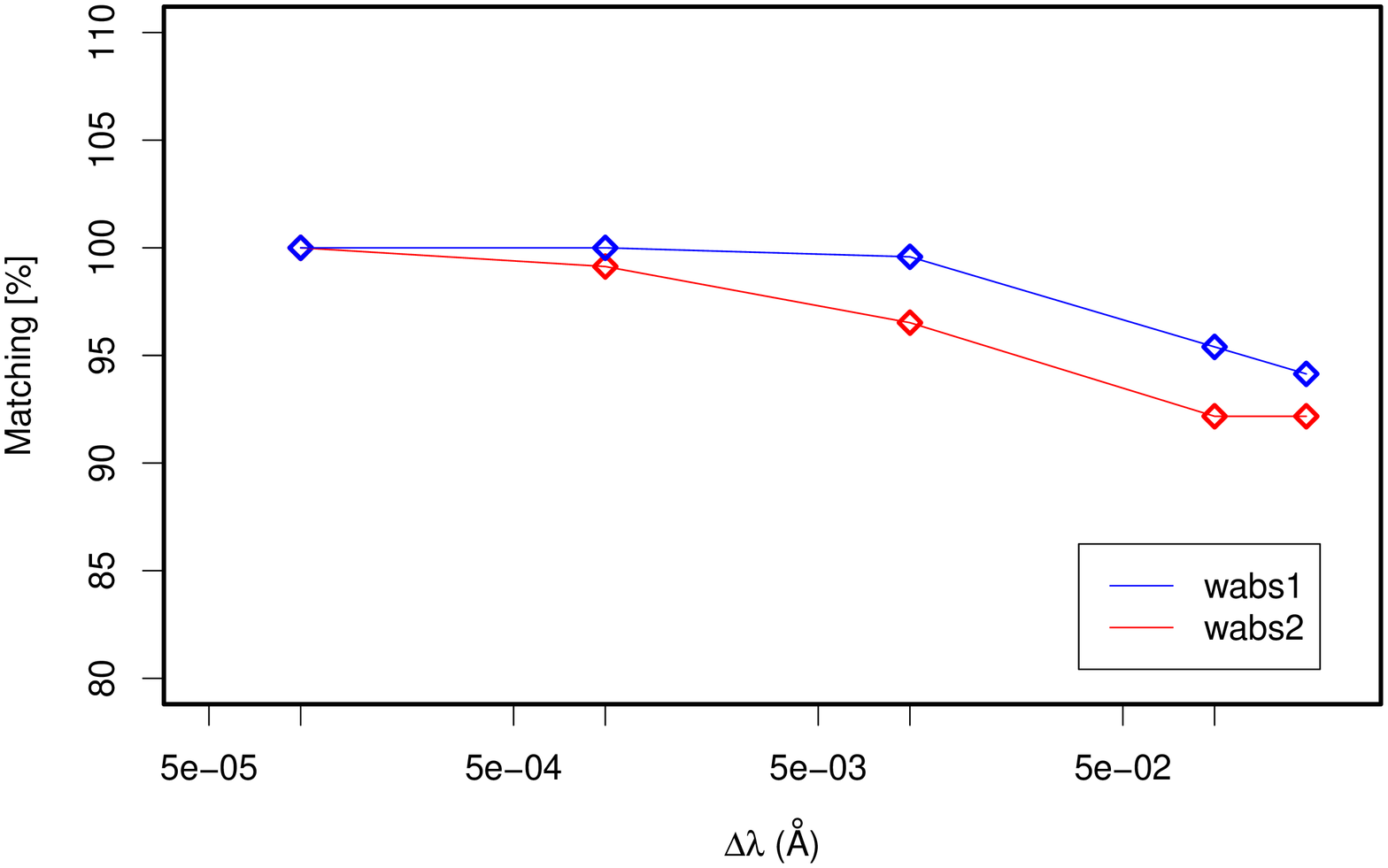}
\caption{
Top panel: Only first stage of the matching lists process
input/{\sc trunk-1}, for the list of the strongest lines
coming from our simulation.
In the X-axis we open
the identification window ($ \Delta \lambda $) from $ 10^{-4}$ to 0.2 \AA.
In blue we present matching process for wabs1,
and in  red for wabs2.
Bottom panel: Second stage of the matching lists process input/{\sc trunk-1}.
In the X-axis we open
the identification window ($ \Delta \lambda $)
from $10^{-4}$ to 0.2 \AA. In blue we present matching process for wabs1,
and in  red for wabs2 (see Y-axis scale).
}
\label{matching_lista1}
\end{figure}
To demonstrate the efficacy of our methodology,
we have applied the {\sc trunk-1} algorithm to the synthetic spectra produced
with Model~A, i.e., the model consisting of two warm absorbers (Table~1).

Because we know exactly which lines are contained in the synthetic spectra,
we can then assess how well {\sc trunk-1} identify the observed lines.
Figure~\ref{matching_lista1} 
shows the level of matching in percent achieved by {\sc trunk-1}.
The upper panel in Figure~\ref{matching_lista1} 
shows the level of matching achieved when
the identification is performed solely based on the line position. A good
matching is only possible when the identification window is set to the 
smallest value ($\Delta\lambda_0=10^{-4}$ \AA). For wider windows the quality
of the identification becomes rapidly poor. A much better performance is 
achieved when, in addition to line positions, line strengths are also compared
with the theoretical predictions. This agreement is shown in the lower panel 
of Figure~\ref{matching_lista1}, 
where is clear that the full {\sc trunk-1} method is capable to 
correctly identify no less of 90\% of the lines, even for the largest 
identification window considered ($\Delta\lambda_0=0.2$ \AA).

The synthetic spectra discussed above contains 239 absorption lines produced
with the MIP (wabs~1), and 115 lines produced with the LIP (wabs~2). Although
the identification is somewhat better for the lines produced in the absorption 
component with the smaller redshift (wabs~1), these results indicate that our 
algorithm can be successfully implemented to real X-ray observations where complex 
mixture of absorption components can be present. Figure \ref{isolated_L1} shows
the 11--12~\AA\ band of the synthetic spectra, where the known absorption features
and those identified by {\sc trunk-1} are indicated. 

Our method correctly identifies
100\% of the lines in this band, which we have tested for different widths of
the identification window. A similar analysis is performed using a synthetic 
spectra based on Model~B (Table~2), 
where on top of the two warm absorber components,
a third component due to galactic absorption is included (ism). This latter component
only includes the continuum (bound-free) photoelectric absorption, thus while it 
modifies the model continuum it does not introduces any additional absorption 
features.

\begin{figure}
\centering
\includegraphics[width=8.5cm]{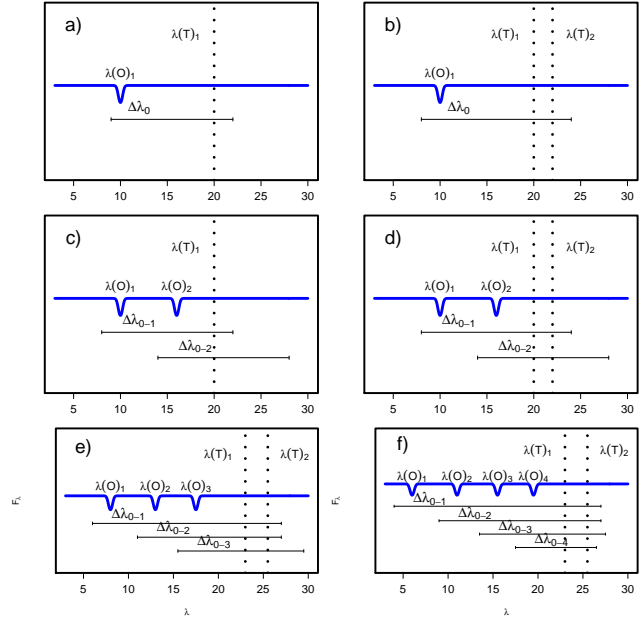}
\caption{
Complex cases treated by {\sc trunk-1} for the ID of lines.
Panel (a) the simplest case.
Panel (b) two candidate lines case.
Panel (c) two observed lines have one theoretical line
as candidate.
Panel (d) two- features, two- candidates (duplets).
Panel (e) two velocity-component case.
Panel (f) two velocity-component duplets case.
See text for details. 
}
\label{cases1}
\end{figure}

\begin{figure}
\centering
\includegraphics[angle=-90,width=8.5cm]{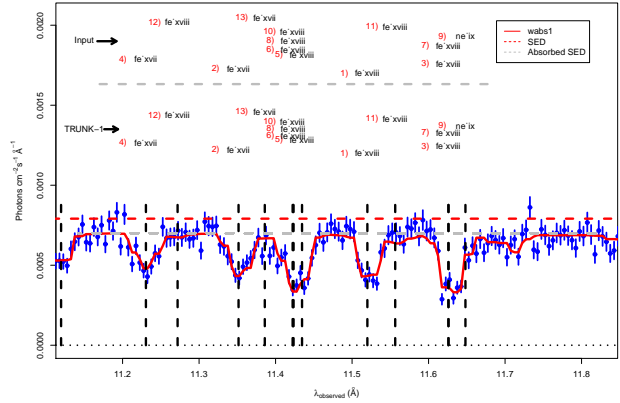}
\caption{
Simulated spectrum (11-12~\AA) showing the input ``full-list" given to the procedure
{\sc trunk-1} for wabs1, and the output identification list, using
$ \Delta \lambda = 10^{-4}$ \AA \sp
(vertical black dashed lines for wabs1). The
absorption lines are enumerated in increasing strength order from 1 to 13.
}
\label{isolated_L1}
\end{figure}

In order to illustrate the virtues and caveats of the method we show
in Figure \ref{cases1}, six (6) possible cases which are easily
solved by it.
In panel (a) we have the simplest possible case. One observed
feature $\lambda(O)_1$, has an IW $\Delta \lambda_0$,
and only the theoretical line $\lambda(T)_1$ is within the IW.
Then $\lambda(O)_1$ is easily identified as $\lambda(T)_1$.

In panel (b), theoretical lines $\lambda(T)_1$ and $\lambda(T)_2$
are within $\Delta \lambda_0$. In this case we simply take the one
with the largest optical depth. If we assign $\tau_1$ to $\lambda(T)_1$
and $\tau_2$ to $\lambda(T)_2$ with $\tau_1 > \tau_2$,
then $\lambda(O)_1$ is simply identified as $\lambda(T)_1$.

In panel (c) we now have a second observed feature $\lambda(O)_2$.
Assuming $\lambda(O)_1$ was already identified as $\lambda(T)_1$ and
$\Delta \lambda_{0-2}$ does not contain another theoretical line,
$\lambda(O)_2$ will be marked as {\sc unknown}. Although this is a rare
case, it can be mentioned as a caveat of the algorithm (which also happens
with regular ID methods).

In panel (d) a second theoretical line $\lambda(T)_2$ is now in the list
of candidates for $\lambda(O)_2$. Assuming $\lambda(O)_1$ was already identified
as $\lambda(T)_1$, the {\sc trunk-1} will force to $\lambda(O)_2$ to be identified
as $\lambda(T)_2$ and not as $\lambda(T)_1$, missing the ``unlikely" possibility
of $\lambda(O)_2$ being a second velocity-component of $\lambda(O)_1$, and
by some unknown mechanism $\lambda(T)_2$ was remove from our line-of-sight.
Although rare, this also has to be mentioned as a caveat of the method.

In panel (e), we allow the possibility of $\lambda(O)_1$ to have a second 
velocity-component, if $\lambda(O)_1$ was already identified as $\lambda(T)_1$,
and $\lambda(O)_2$ as $\lambda(T)_2$, then $\lambda(O)_3$ will be identified
as $\lambda(T)_1$.

And finally in panel (f), there will be two duplets velocity-components with
$\lambda(O)_1$ and $\lambda(O)_3$ identified as $\lambda(T)_1$, and
$\lambda(O)_2$ and $\lambda(O)_4$ identified as $\lambda(T)_2$.

Although not complete, the list of possible cases cover the situations given
in the present work. More cases and/or possibilities can be added to the
algorithm in the future, but the results given by Figure 3 (bottom panel),
show the method's performance is at the level of $\approx 90$ \%.

In summary we list the limitations of the method:
\begin{enumerate}

\item From the beginning, if the photoionization components
(e.g., wabs1 and/or wabs2) are not globally fitting the spectrum,
the ID process will be dubious.

\item The method expects a list of detected features. If
as part of a blend, only
one feature is given, the method will ID it as the strongest one,
regardless the micro-physics of the blend.

\item The method will not identify a case where the physics changes
the ratio $\lambda(T)_1/\lambda(T)_2 > 1$ to $<1$ (e.g., case b of
Figure \ref{cases1}).

\item Naturally, if the continuum is not well established in the building
of the physical model, at the beginning of the process, the
absolute measurement of EWs will be wrong (though the comparison between
them will be good since they all use the same continuum).

\item The algorithm does not allow the ID of
two-velocity component of a feature under the
configuration given by case c) of Figure \ref{cases1}.
In all those cases visual and manual inspections are
required.

\end{enumerate}


\subsection*{Line Profile and EWs}

Our procedure receive a list of lines 
(previously produced in the ID step)
and applies the following treads:
Each transition produces an absorption optical depth computed with:
\begin{equation}
\tau_{\lambda}=\tau_{0}\phi_{\lambda},
\end{equation}
\noindent
the optical depth at the core of the line is
$\tau_{0}=\int^{\Delta R}_{0}\tau_0(r) dr = 
N_{\rm ion}^0\frac{\pi e^2}{mc}f_{\rm lu}$,
where $\Delta R=N_H/n_H$ is the thickness of the
slab,
$N_{\rm ion}^0$ is the column density of the ion at the line core
(i.e., $\lambda=\lambda_0$),
$m$ is the electron mass,
$e$ is the electron charge, $c$ is the speed of light,
and
$f_{\rm lu}$ is the oscillator strength of the transition.
The line profile $\phi_{\lambda}$,
is assumed to be a Voigt profile
(\cite{rybicki1979a}). The Doppler broadening parameter
$v_{\rm Dop}^2=v_{\rm turb}^2+v_{\rm therm}^2$, 
is the combination of microturbulent ($v_{\rm turb}$) and thermal broadening ($v_{\rm therm}$). 
The last one is computed
with $v_{\rm therm}=\sqrt{\frac{2k_{\rm B}T}{\mu m_H}}$, where $k_{\rm B}$ is the
Bolztmann constant, $\mu m_H$ is the average mass per particle; $\mu\simeq 0.6$ for fully
ionized solar-metallicity gas.
The temperature $T$ is taken
from our \xstar \sp calculations.
The microturbulent velocity $v_{\rm turb}$ is set to 200 \kms.
Then, we build absorption spectrum for each line ($F_{\lambda}$ around each feature)
and compute
EWs:
\begin{equation}
EW=\int{1-\exp{(-\tau_{\lambda})}}d\lambda.
\end{equation}

\begin{figure}
\centering
\includegraphics[angle=-90,width=8.5cm]{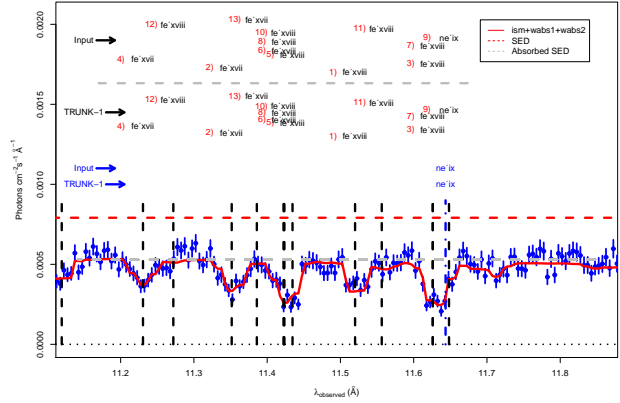}
\caption{
Simulated spectrum (11-12~\AA) showing the input ``full-list" given to the procedure
trunk-1 for ism+wabs1+wabs2, and the output identification list, 
using $ \Delta \lambda = 10^ {-4}$ \AA \sp
(vertical black dashed lines for wabs1, and
vertical blue dashed lines for wabs2).
The
absorption lines are enumerated in increasing order from 1 to 13.
}
\label{isolated_L2_tbabs}
\end{figure}

One important problem we are able to address here, is the difference of
measuring absorption line EWs taking into account all the relevant
atomic physics, i.e., bound-bound and bound-free transitions, which
extend from the SED, in contrast with measurements of EWs made using
as the continuum the absorbed SED.
We call the first type unabsorbed EWs and the second one absorbed EWs.
Figure \ref{isolated_L2_tbabs}, depicts
this problem in a visual manner.

For instance in a model where a single absorber is present (wabs1),
we find difference between the level of SED (red dashed line in Figures),
and the level of the absorbed SED (gray dashed line), which is the result
of the SED's radiation passing through wabs1. It is clear that measuring EWs using
these two different continuum levels will give different results.


Likewise applied to a real observation, the method identifies and present
results from H- and He-like ions falling in the X-rays regime, and differences
among 6 \% at $\lambda \approx 20$ \AA \sp in lines like Ly$_\alpha$ and
Ly$_\beta$ of OVIII, to 24 \% at $\lambda \approx 10$ \AA \sp in lines
like MgXI and L-shell of FeXVI-FeXIX, are attributable to the chosen
continuum level.

The situation becomes more dramatic as the complexity in absorption
increases for model~A and model~B.
We quantify
how important this continuum level problem results, where
ISM is taking into account.
Up to 80 \% differences, in both absorbers wabs1 and wabs2, of
EW
measurements are
observed, and it
serves as benchmark for comparison of EWs measured using
the free-line region (FLR) technique, which make use of absorbed SED
as continuum level,
as for instance in \cite{kaspi2002a}, for measuring line properties of
NGC 3783 (see their Table 1).


\section*{Results: Deducing distance from the supermassive black hole}
\label{resul1}

In this section we apply our method to
the X-ray 900 ks spectrum, introduced in
the first section.

\subsection*{Measured {\it vs} Theoretical EWs}

\begin{figure}
\centering
\includegraphics[angle=-90,width=7.5cm]{fig10.eps}
\includegraphics[angle=-90,width=7.5cm]{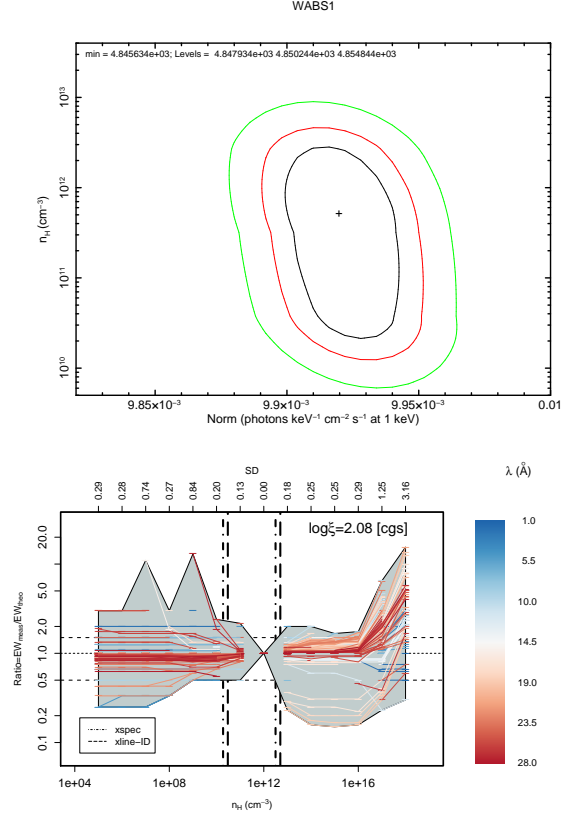}
\caption{
Top panel: 63 \%, 90 \% and 95 \% 2D confidence level
contours on the gas density, $ n_H $, and
normalization parameter infer for wabs1.
Bottom panel: Ratio=$ \frac{EW_{meas}}{EW_{theo}} $ vs $ n_H $,
covering 14 order of magnitudes. At the right side there is a wavelength-color
map for the 239 spectral lines detected for the wabs1.
Top axis presents the standard deviation (SD) from the mean 1,
for each set of lines given by each $ n_H $.
Vertical dot-dashed lines are lower- and upper
limits computed using spectral fitting within
xspec.
Vertical dashed lines are 
lower- and upper
limits computed using 
xline-ID.
}
\label{den_dist2}
\end{figure}

Theoretical curves of
the Ratio=$\frac{{\rm EW}_{\rm meas}}{{\rm EW}_{\rm theo}}$
(measured EW to theoretical EW)
{\it vs}
$n_H$,
for each of the identified features included in the
simulation, are compared for wabs1 and wabs2.
In Figure \ref{den_dist2} (bottom panel) we plot
all the Ratios produced through identification of lines
(239) using the {\sc trunk-1} model
and the absorber wabs1 ($\log \xi=2.08$, $N_H=10^{22}$ \cmn, $n_H=10^{12}$ \cmd).
A wavelength-color map has been placed at the right side, so one
can read which lines are causing the deviations from 1, 
that is obtained when ${\rm EW}_{\rm meas}={\rm EW}_{\rm theo}$.
It is natural from this plot to infer what is the spectrum that
matches better the observation, i.e., the one with the smaller
standard deviation (SD, top axis on the plot) from
the mean equal to 1.
The same type of graphics has been plotted in Figure \ref{den_dist1}
for wabs2 ($\log \xi=1.15$, $N_H=10^{21}$ \cmn, $n_H=10^{12}$ \cmd),
where again we reach the conclusion that the best-fit
$n_H$ is the one with smallest SD for Ratio, with mean 1.
An analysis of these results will be given in the next section.
\subsection*{COG analysis}

\begin{figure}
\centering
\includegraphics[angle=-90,width=7.5cm]{fig12.eps}
\includegraphics[angle=-90,width=7.5cm]{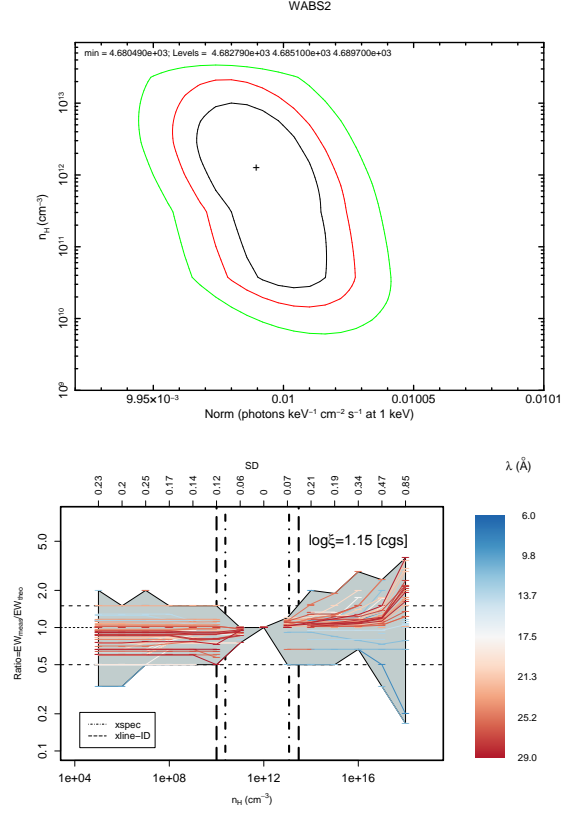}
\caption{
Top panel: 63 \%, 90 \% and 95 \% 2D confidence level
contours on the gas density, $ n_H $, and
normalization parameter infer for wabs2.
Bottom panel: Ratio=$ \frac{EW_{meas}}{EW_{theo}} $ vs $ n_H $,
covering 14 order of magnitudes. At the right side there is a wavelength-color
map for the 115 spectral lines detected for the wabs2.
Top axis presents the standard deviation (SD) from the mean 1,
for each set of lines given by each $ n_H $.
Vertical dot-dashed lines are lower- and upper
limits computed using spectral fitting within
xspec.
Vertical dashed lines are
lower- and upper
limits computed using
xline-ID.
}
\label{den_dist1}
\end{figure}

In order for a line to be used as density diagnostic,
it must be not saturated, and its ratio with measured
EWs sensitive respect to $n_H$.
In Figure \ref{den_dist1} we present the ratio
of measured to theoretical EWs for each of
the feature identified and included in building
of the spectrum, given by the LIP, with
IW, $\Delta \lambda = 10^{-4}$ \AA.
This plot infers that the plasma gas density reproducing
best the observation is
$n_H=10^{12}$ \cmd, where SD=0.0. Fitting procedures allow us
to set limits on SD in order to achieve confidence level in therms
of $\chi^2$. SD of 0.12-0.15 produce changes $> 90$ \% confidence
level on measurements of $n_H$.
We are able to infer the gas density of the spectrum to be
$n_H=(2.4\times 10^{10}, 1.2\times 10^{13})$ \cmd, for the LIP
component of the absorbing complex in model~B.

The COG ratio shown by the MIP is an excellent case of highly sensitive
curve with $n_H$ (see Figure \ref{den_dist2}).
The envolvent (irregular gray polygon)
makes a strong constrain of the gas density to be
$n_H\approx 10^{12}$ \cmd.
Again limits on the measurements using
SD=0.15-0.17 ($\chi^2 > 2.79$)
reports $n_H=(1.9\times 10^{10}, 3.1\times 10^{12})$ \cmd,
for the MIP component of model~B.

We checked for the robustness of these measurements
with the building of a 2D contours, produced by
computing a grid of photoionization models for wabs1
(MIP) and wabs2 (LIP), varying $\log (n_H)=5-18$ [\cmd],
with one point per decade. The result is plotted at the top
panel of Figure \ref{den_dist2} (for wabs1).
From the best-fit
$n_H\approx 10^{12}$ \cmd, $\chi^2\approx 4845$
using 4604 PHA bins (and two free parameters), we
compute contour level of confidences
63 \%, 90 \% and 95 \% for $n_H$ and the
normalization parameter of the powerlaw.

We remark here, that in an independent way, we are obtaining
results that are in agreement with the measurements given by our own method.
In the bottom panel we compare these {\sc xspec} measurements,
i.e., single parameter 90 \% confidence level on $n_H$
(vertical dot-dashed line),
against the {\sc xline-ID} results.
The obvious advantage of our method,
is that we can track each of the absorption line which is
contributing to the COG ratio density diagnostic,
overcoming problems like several $\chi^2-$minima,
double evaluated best-fit parameter,
lack of ability to evaluate what are the spectral ranges contributing
to $\chi^2$, among others.

\subsection*{Velocity density distribution}
\label{veldist1}

\begin{figure}
\centering
\includegraphics[angle=-90,width=8.5cm]{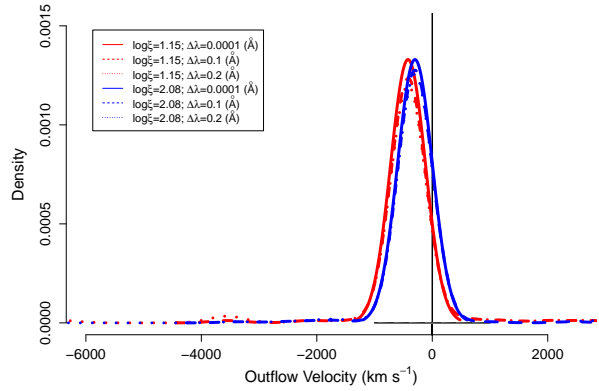}
\caption{
Probability distribution function (p.d.f) for the flowing velocity of the material
absorbing X-rays in the 900 ks \chan \sp simulated spectrum.
Solid, dashed and dotted lines are smoothed p.d.fs
for the 
$ \Delta \lambda $= $ 10^{-4}, 0.1, 0.2$
respectively. It is used a Gaussian kernel with the best bandwidth
of $ h=300 $ \kms.
Red and blue are LIP and MIP plasmas
respectively.
}
\label{vel_dist1}
\end{figure}

The second main
result of the present
analysis can
be drawn from Figure \ref{vel_dist1}.
There we plot the 
univariate kernel density estimators (k.d.e) for the
Doppler
velocity distribution:
\begin{equation}
\hat{f}_{\rm kern}(v,h)=\frac{1}{nh}\sum_{i=1}^n{K \left( \frac{v-V_i}{h} \right) },
\end{equation}
where the kernel function $K$ is a Gaussian and normalized to unity, 
i.e., $\int K(x)dx=1$,
$h$ is the (constant) bandwidth
set to 300 \kms, and $V_1,V_2,...V_n$
is our random sample of velocities of size $n$.
Each possible identification list has the
500 strongest lines for each point of the grid and, as 
mentioned in section {\bf Models: The grid}, there are 28 grid points.
Whenever a detected feature is out of the identification window
($\Delta \lambda$ \AA \sp around the feature),
is flagged as {\sc unknown} and is not taken into account in the 
probability distribution function
(p.d.f)
analysis. 
As we may observe
from Fig. \ref{vel_dist1}, one of the main components have peaks at
$v_{\rm out}\approx 300~({\tt wabs1})$ and 420~({\tt wabs2})
\kms,
confirming the expected result,
as this is what we introduce as outflow velocity
in our simulated data (see Table~2).

A closer examination also reveals
densities different from zero for
higher velocity outflows, 
for instance exhibits by the LIP 
Si~{IX} $\lambda 6.938$ \r{A}
outflowing at $v_{\rm out} \simeq 3600$
\kms, and 
Si~{VIII} $\lambda 7.005$ \r{A}
with $v_{\rm out} \simeq 3200$ \kms,
which are clearly wrong identifications,
consequence
of the identification window scheme
(for the two plasma,
solid line has $\Delta \lambda = 10^{-4}$ \AA,
dashed line with $\Delta \lambda = 0.1$ \AA,
and dotted line with $\Delta \lambda = 0.2$ \AA).
However we see that the density function of these
velocities are very small compared with the components
made by right identification.

\section{Discussion \& Conclusions}
\label{discu1}
In this
paper
we show
how important is to correctly define the level of continuum
for EW measurements, which could differ up to $\sim 80$ \%,
including or not interstellar absorption.
We also show that {\sc xline-id} is better than 90\% in the identification
of absorption lines, where more than one absorber is present.

\begin{figure}
\centering
\includegraphics[angle=-90,width=8.5cm]{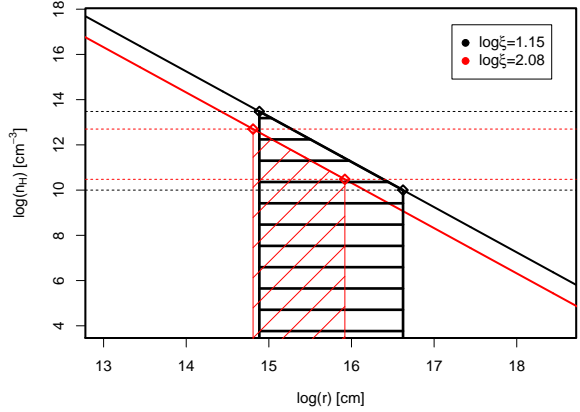}
\caption{
Spatial location of the absorbers in the analyzed data. The gas density $n_H$ is
independently inferred from our COG ratio method. Horizontal lines: MIP (red) and LIP (black).
Vertical lines are lower and upper limits on $ r $ deduced from equation (\ref{xi11}).
}
\label{distance1}
\end{figure}

Once we have inferred the gas density from the spectral
lines, we can take equation (\ref{xi11}) and compute
$r$ from it. What is deduced,
using the relation $n_H  \propto r^{-2} $, and Figure
\ref{distance1}, is a suggestion of the MIP located at distances
of $r_{MIP}\simeq (0.18-1.80)$ lt-days 
($\log r \simeq 14.7-15.7$ [cm]),
and
the LIP located at distances 
of $r_{LIP}\simeq (0.51-5.13)$ lt-days 
($\log r\simeq 15.1-16.1$ [cm]).
The global picture that emerges from this is that highly
($\log \xi \sim 2$) ionized material with gas density
($n_H \sim 10^{12}$ \cmd) is located closer than the less
ionized (colder $\log T\sim 4.7$ K) gas.
There was no attempt to introduce any pre-established density
profile or velocity law in the procedure, and the final relationship
between $n_H$ and $r$ was self-consistently achieved.

\textsf{acknowledgments}

This work is partially supported by IVIC project 2013000259.
Also, it was partially supported by ABACUS, CONACyT (Mexico)
grant EDOMEX-2011-C01-165873.

\end{document}